\documentclass[%
 aip, apl,
 amsmath,amssymb,
 reprint,%
]{revtex4-1}

\usepackage{graphicx}
\usepackage{dcolumn}
\usepackage{bm}

\usepackage[utf8]{inputenc}
\usepackage[T1]{fontenc}
\usepackage{mathptmx}

\begin{document}

\preprint{AIP/123-QED}

\title[]{ \emph{Ab initio} study of band gap properties in novel metastable BC8/ST12 Si$_x$Ge$_{1-x}$ alloys}

\author{J. Wagner}
 \email{jowagner@gfz-potsdam.de}
\affiliation{ 
Helmholtz-Centre Potsdam GFZ, German Research Centre for Geosciences, Telegrafenberg D-14473 Potsdam, Germany
}%
\author{M. N\'u\~nez-Valdez}%
 \email{mari_nv@gfz-potsdam.de}
\affiliation{ 
Helmholtz-Centre Potsdam GFZ, German Research Centre for Geosciences, Telegrafenberg D-14473 Potsdam, Germany
}%
\affiliation{Goethe University Frankfurt am Main, Altenh\"oferallee 1, 60438 Frankfurt am Main} 

\date{\today}

\begin{abstract}
The cubic $Ia\overline{3}$ (BC8) and tetragonal $P4_32_12$ (ST12) high pressure modifications of Si and Ge are attractive candidates for applications in optoelectronic, thermoelectric or plasmonic devices. 
Si$_x$Ge$_{1-x}$ alloys in BC8/ST12 modifications could help overcome the indirect and narrow band gaps of the pure phases and enable tailoring for specific use-cases. Such alloys have experimentally been found to be stable at ambient conditions after release from high pressure synthesis, however their fundamental properties are not known. In this work, we employ {\it ab initio} calculations based on density functional theory (DFT) to investigate the electronic properties of these compounds as a function of composition $x$. We obtain the effective band structures of intermediate alloys by constructing special quasi random structures (SQS) and unfolding their band structure to the corresponding primitive cell. Furthermore, we show that the indirect band gap of the ST12 Ge end-member can be tuned to become direct at $x_\text{Si} \approx 0.16$. Finally, our investigations also demonstrate that the BC8 modification, on the other hand, is insensitive to compositional changes and is a narrow direct band gap semiconductor only in the case of pure Si.
\end{abstract}

\maketitle
High pressure modifications of silicon and germanium have seen substantial research in the last decades as many of these modifications can be stabilized at ambient conditions and exhibit properties highly sought after in materials design \citep{hab16}. The cubic BC8 and tetragonal ST12 modifications are no exception and have seen a number of studies investigating their electronic structure, optical- and thermoelectric properties \cite{mal08,wip13,hab16}. Both modifications are obtained by pressure release at about 10 - 12 GPa from their respective $\beta$-Sn modification \cite{kas64,muj93}. Both Ge and Si can be realized in metastable BC8 structures \cite{nel93, muj03}, but only Ge has been synthesized in the ST12 modification \cite{kas64}. Although, there has been some debate concerning the electronic structure of ST12 Ge. Early studies suggest a direct fundamental band gap, thus promising a viable alternative for solar absorber applications as compared to common diamond cubic (DC) silicon \cite{muj93}. However, more recent experimental and theoretical studies indicate that the ST12 Ge band gap is, in fact, slightly indirect \cite{mal12a, zha17a}, yet smaller than the indirect band gap of DC Ge. BC8 Si has a similar history. While former studies found the BC8 Si modification to be a semimetal \cite{kas64,mal08}, a recent experimental work suggested a narrow gap semiconductor ($\sim$35 meV) exhibiting reduced thermal conductivity and potential for laser applications \cite{zha17b}. 
In order to optimize any of these properties, it is worth looking at similarities to other group IV systems such as Ge$_x$Sn$_{x-1}$, where alloying DC Ge with relative small amounts of Sn leads to a well known transition from indirect to direct band gap \cite{wir15}. A similar procedure for ST12 Ge could lead to a direct band gap material with high Si content and some potential for today's chiefly Si based semiconductor industry. The same holds true for a potential composition-tunable band gap in the BC8 Si. Thus, Si$_x$Ge$_{1-x}$ alloys in both the BC8 and ST12 modifications have been synthesized in the entire compositional range $0\le x\le 1$ \cite{ser14}. It was found that upon pressure release, the ST12 structure is retained at least up to $x\approx0.25$, whereas for $x>0.25$, BC8 is formed. Other than this experimental study, to the best of our knowledge, there are no reports on the atomic structure or the electronic properties of these alloys as a function of $x$. To address this issue, we performed state-of-the-art density functional theory simulations of bulk Si$_x$Ge$_{1-x}$ BC8/ST12 alloys using both primitive cell (PC) and supercell (SC) calculations. We rigorously modeled all atomic configurations for selected compositions and investigated how atomic arrangements affect the band gap and phase stability. 
\begin{figure}
\includegraphics[width=0.49\textwidth]{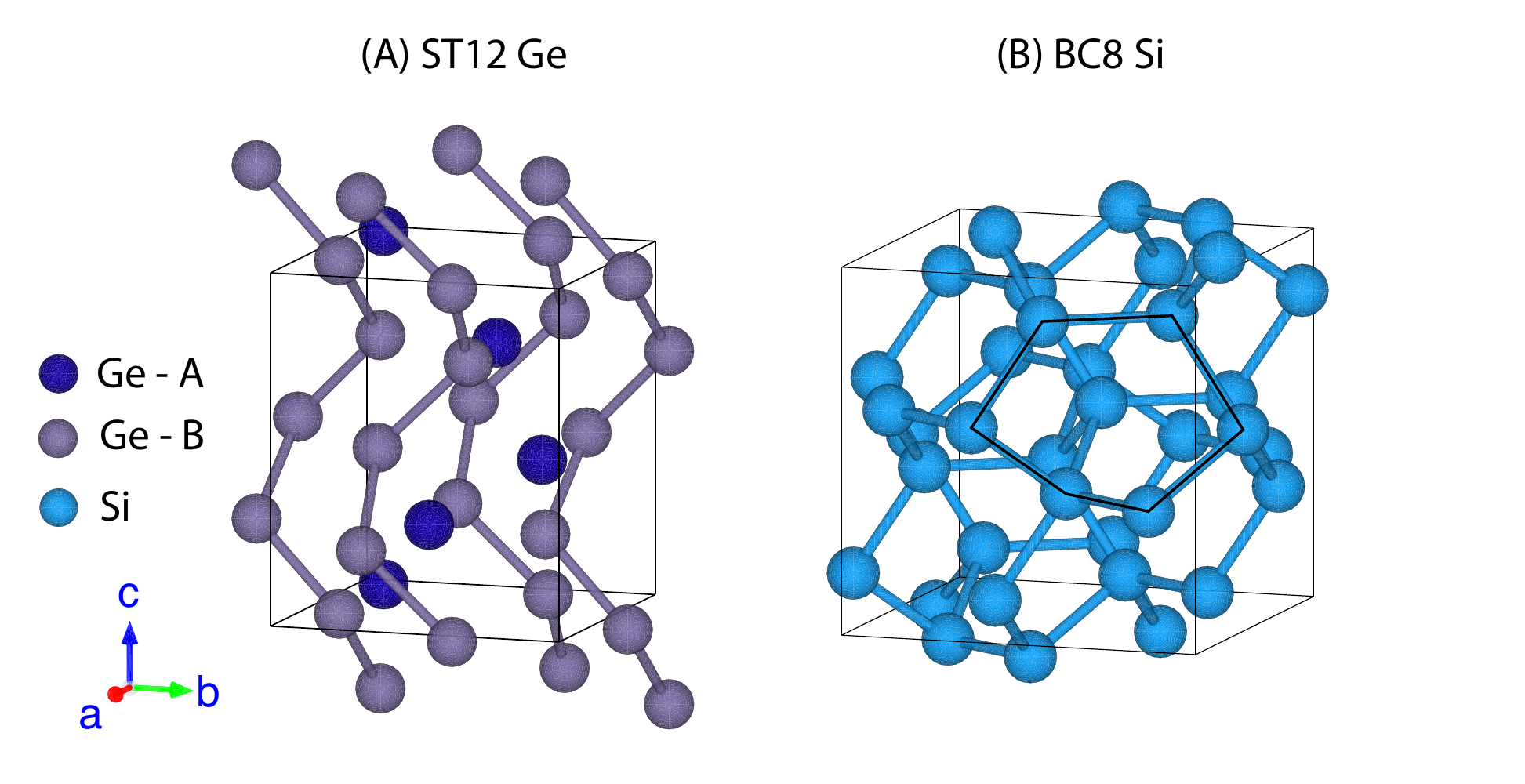}
\caption{Conventional unit-cells of the ST12 Ge and BC8 Si end-members. (A) In the ST12 Ge structure, additional atoms along the $c$-axis have been added to illustrate the spiral-chain arrangements formed by atoms in Wyckoff position B. Bonds to A atoms are not shown for simplification. (B) BC8 Si: The 6 fold ring arrangement in the structure is shown by the thin black line (see text).}\label{fig:struct}
\end{figure}
Both the ST12 and BC8 structure have been extensively described by various authors elsewhere \citep{kas64, nel93, nee95, muj93, mal08, mal12a}. The tetragonal ST12 structure with space group $P4_32_12(D^4_8)$, has a 12-atom unit cell with two Wyckoff positions. Four atoms on position A and eight on position B. Atoms of type B form fourfold spiral chains, propagating parallel to the conventional $c$-axis (see Figure \ref{fig:struct}a). These spirals are linked by tetrahedra with an atom of type A in the center and a type B atom belonging to separate chains at each corner. A possible formation pathway from $\beta$-Sn $\rightarrow$ ST12 is a local bond twisting mechanism \citep{wan13}. This process results in strong deviations in tetrahedral bond angles as compared to the DC structure tetrahedron, while bond distances are preserved.\\
The BC8 structure (Figure \ref{fig:struct}b), spacegroup $Ia\overline{3}(T^7_h)$, has eight atoms in its primitive unit cell, forming a body-centered-cubic lattice. All atomic sites are symmetrically equivalent and the structure is fully described by its lattice constant and one internal parameter. In BC8, bond distances are distorted as compared to the respective DC modification while bond angles are more or less preserved. Thus, the structure can be viewed as an arrangement of highly distorted six-fold rings.\\
In this work, we chose a combined approach to model the Si$_x$Ge$_{1-x}$ BC8 and ST12 alloys. Firstly, to analyze order/disorder effects on phase stability and lattice parameters, we modeled all possible atomic site occupancies in the primitive cell of ST12 and BC8, using the package \textit{Site-Occupation-Disorder} (SOD) \citep{gra07}. By taking into account space group symmetry, we modeled a total of 362 configurations for $ x = 0, 0.08, 0.16, 0.25, 0.33, 0.5, 0.66, 0.75, 0.84, 0.92, 1$ in the ST12 modification, and a total of 475 configurations for $x = 0, 0.06, 0.125, 0.18, 0.25, 0.5, 0.75, 0.875, 1$ in the BC8 modification (eight and 16 atoms/cell). Secondly, we generated randomly distributed supercells for all intermediate compositions, as the experimentally synthesized alloys are expected to be fully disordered \cite{ser14}. We achieved this by generating special quasi random structures (SQS) \cite{zun90} of $2\times2\times2$ supercells using the SQS algorithm implemented in the {USPEX} package \cite{gla06,lya13}. 

For all structural relaxations, we performed \textit{ab initio} calculations using the projector augmented wave (PAW) approach \citep{blo94}. For the exchange correlation energy we employed the general gradient approximation in the revised Perdew-Burke-Ernzerhof (PBEsol) \citep{per08} formalism as implemented in the VASP code \citep{kre94, kre96}. 
Notice that we also employed standard PBE and LDA calculations to reproduce previously reported literature data. As they performed identical to published data, we chose not to include them and focus on our PBEsol results, which reproduce the available experimental data much better (see below). 
As a compromise between efficiency and accuracy, a planewave cutoff of 500~eV was chosen. Full structural relaxation of cell volume and atomic positions for all configurations were performed with 10$^{-3}$~eV and 10$^{-8}$~eV convergence criteria in force and energy, respectively. To keep the results comparable across different compositions, we chose a consistent $\Gamma$-centered k-point mesh with a 0.2~\AA$^{-1}$ sampling rate. For density of states (DOS) and band structure calculations, we decreased the spacing to at least 0.07~\AA$^{-1}$. These criteria led to $6 \times6\times6$ and $16\times16\times16$ k-point meshes for structural relaxation and DOS calculations, respectively. For the much larger SQS, the respective meshes had to be reduced to $4\times4\times4$ and $9\times9\times9$.
Lastly, we investigated the electronic band structure of end-members and intermediate compositions represented by the SQS. Notice, however, that electronic band structures are only well defined for periodic crystals where Bloch's theorem is valid. Naturally, this is not the case for random alloys where space group symmetry is formally broken. To overcome this, we followed the effective band structure approach (EBS)\cite{pop10}. Any SQS supercell is geometrically linked to the corresponding PC by simple lattice vector translations. This enables the unfolding of any SC band structure to the Brillouin zone of the PC. During band unfolding, each state in the PC is assigned a spectral weight that reflects how well that state is preserved in the random SC. In this work, unfolding of selected Si$_x$Ge$_{1-x}$ SQS has been performed using the BandUP code \cite{med14}. Band gap energy ($E_{bg}$) and position were determined using the SUMO code \cite{gan18}.
In a recent study of BC8 Si, simulations in the Heyd-Scuseria-Ernzerhof (HSE) formalism \cite{hey03} were performed to accurately reproduce the measured experimental band gap \cite{zha17b}. In order to better compare to these results, we also performed additional HSE06 calculations to give an estimate of its impact on the $E_{bg}$ values.\\

\begin{figure}
\includegraphics[width=0.49\textwidth]{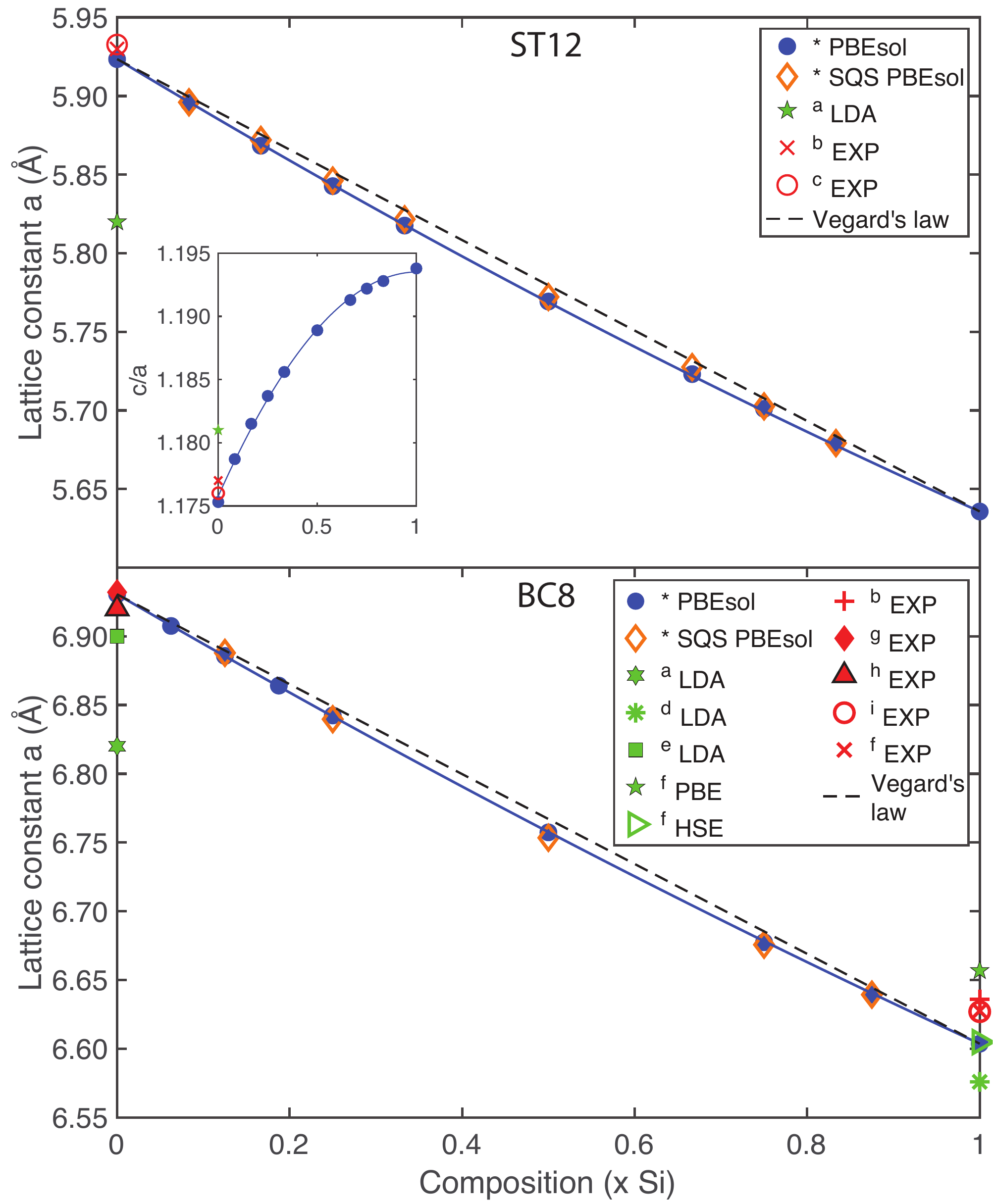}
\caption{Lattice constant $a$ and bowing curve for ST12 (top) and BC8 (bottom) Si$_x$Ge$_{1-x}$ alloys. Blue dots are the averaged value of all possible site occupancy configurations for one individual composition $x$. The inset shows the $c/a$ ratio for all ST12 alloys. Orange diamonds are the lattice parameters directly derived from special quasi-random structures (SQS). Theoretical error bars are smaller than symbols. References for the data are: $[*]$-This work, a-[\onlinecite{mal12a}], b-[\onlinecite{kas64}], c-[\onlinecite{zha17a}], d-[\onlinecite{mal08}], e-[\onlinecite{muj93}], f-[\onlinecite{zha17b}]; g-[\onlinecite{nel93}], h-[\onlinecite{bat65}], and i-[\onlinecite{kur16}].}\label{fig:lattice}
\end{figure}

\begin{table}
\caption{\label{tab:table_struct}
Lattice parameter $a$, ratio $c/a$, and band gap energy ($E_{bg}$) of ST12 and BC8 Si/Ge end-members at zero pressure. Band gap energies refer to the indirect band gap in the ST12 case, and a direct band gap in BC8. [*]: This work, EXP: experimental study, and [$\dagger$]:~Conductivity measurements.}
\begin{ruledtabular}
\begin{tabular}{l|llll|lll}
 &\multicolumn{4}{c|}{ST12} & \multicolumn{3}{c}{BC8} \\
&$a$ (\AA) & $c/a$ & $E_{bg}$(eV) & Reference & $a$(\AA) & $E_{bg}$(eV) & Reference \\
\hline
Ge	& 5.923	& 1.175		& 0.44			& PBEsol$^*$	& 6.931		& metal		& PBEsol$^*$\\
 		& -			& 		-		& 1.01			& HSE06$^*$	& -			& indirect		& HSE06$^*$\\
 		& 5.930	& 1.177		& -				& EXP\cite{kas64}			& 6.932	& -			& EXP\cite{nel93}\\
		& 5.933	& 1.176		& 0.63$^\dagger$	& EXP\cite{zha17a}			& 6.920		& -			& EXP\cite{bat65}\\ 
 		& -      		&  -     		& 0.70		& HSE06\cite{zha17a}  		& 				&				&\\ 
 		& 5.927  	& 1.179 		& 0.70			& LDA\cite{muj93}			& 6.900		& metal		& LDA\cite{muj93}\\ 
 		& 5.82	0	& 1.181		& 0.54			& LDA\cite{mal12a}			& 6.820		& metal		& LDA\cite{mal12a}\\
\hline
Si		& 5.635	& 1.194		& 0.99		& PBEsol$^*$	& 6.604	& metal		& PBEsol$^*$\\
		&				&				&					&		& -			& 0.04		& HSE06$^*$\\
		&				&				&					&		& 6.628	& 0.03		& EXP\cite{zha17b}\\
		&				&				&					&		& 6.605	& 0.01		& HSE06\cite{zha17b}\\
		&				&				&					&		& 6.636	& -			& EXP\cite{kas64}\\
		&				&				&					&		& 6.657	& metal		& PBE\cite{zha17b}\\
		& -			& -			& 1.10				& LDA\cite{mal08}			& 6.576	& metal		& LDA\cite{mal08}\\ 
\end{tabular}
\end{ruledtabular}
\end{table}
Table \ref{tab:table_struct} lists lattice parameters and band gap energies of the Si/Ge pure phases, comparing literature values and results of this work. After relaxation of ST12 Ge, we found the lattice parameter $a = 5.923$~\AA\ and the ratio $c/a = 1.175$, which are in excellent agreement to experimental values, see Table \ref{tab:table_struct}. For ST12 Ge, we found an indirect band gap of 0.436~eV (direct 0.443~eV) with the conduction band minimum (CBM) at [0.35, 0.35, 0.00] and valence band maximum (VBM) at [0.34, 0.34, 0.00]. This is in good agreement with previous DFT studies\cite{mal12a,zha17a}. Calculating the band gap once more using the HSE06 functional with a Fock exchange of 25\%, while keeping the structure from the PBEsol optimization, resulted in an opening of the indirect band gap to 1.006~eV (direct 1.014~eV). For ST12 Si, we predicted its lattice parameter to be $a = 5.635$~\AA, its ratio $c/a = 1.194$, and an indirect band gap of 0.998~eV (direct 1.348~eV), see Table \ref{tab:table_struct}, with the VMB located at [0.32, 0.32, 0.00] and the CBM just of the Z-point [0.00, 0.00, 0.49]. These results are in qualitative agreement with the previous DFT study which employed a much denser k-mesh\cite{mal08}.
After structural relaxation of the BC8 phases, we found a lattice parameter of $a = 6.931$~\AA\ for pure Ge, which is in excellent agreement to the experimentally determined values \cite{nel93, bat65}. The lattice parameter of BC8 Si is slightly off the experimental value but within reasonable agreement compared to other PBE/LDA studies, as can be seen clearly from Table \ref{tab:table_struct} and Figure \ref{fig:lattice}. Using PBEsol for the band structure, we predicted a metallic nature with a small overlap for both Si and Ge end-members. In the special case of BC8 Si, band structures based on HSE06 calculations with 35\% Fock exchange resulted in an opening of the band gap to 0.043~eV, which is surprisingly consistent with the value determined by a combined study of experiment and HSE06 calculations \cite{zha17b}. On the other hand, BC8 Ge shows an indirect band gap of 0.003~eV (direct 0.187~eV) upon HSE06 calculation with the same parameters.

\begin{figure}
\includegraphics[width=0.49\textwidth]{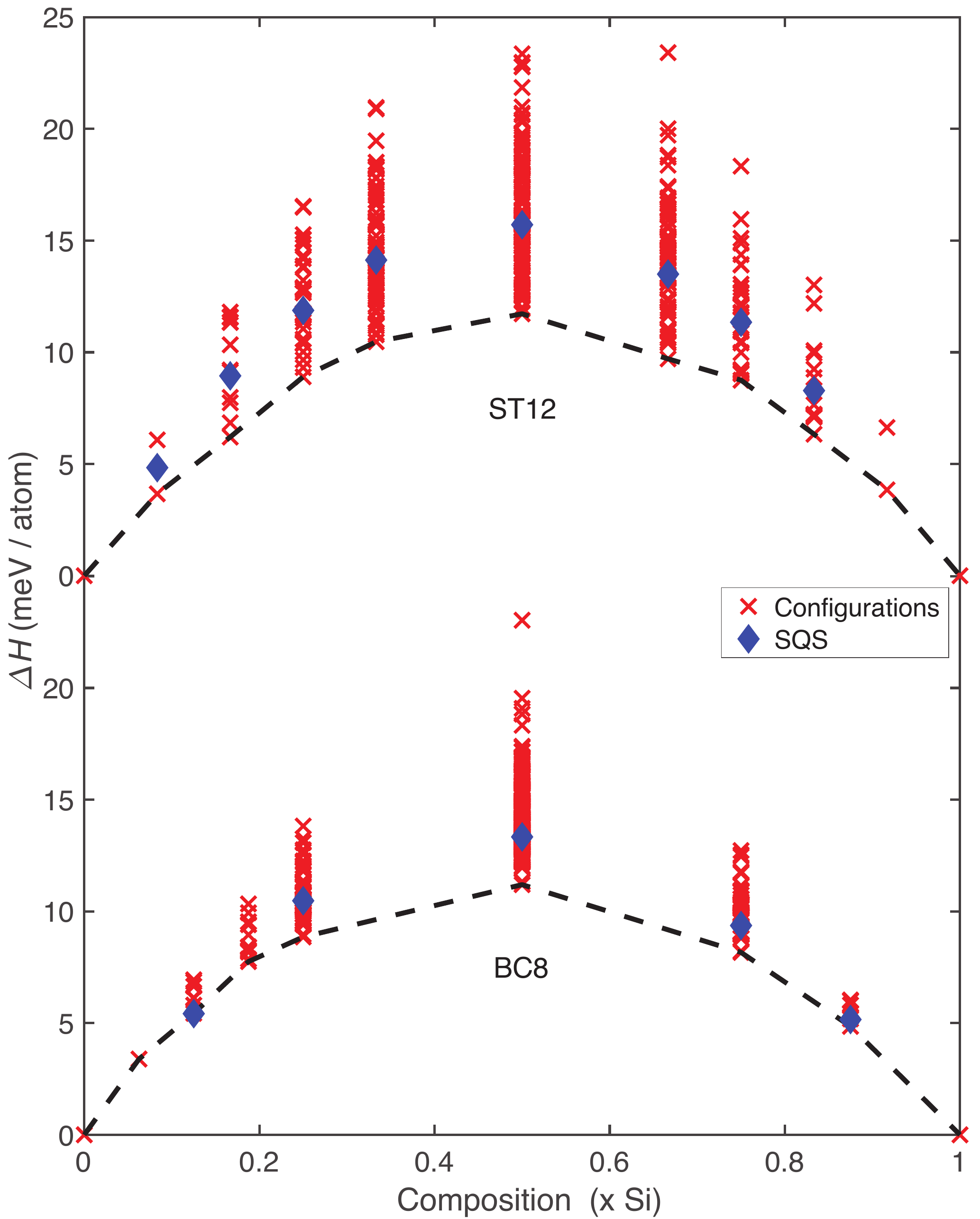}
\caption{Convex hull of ST12 and BC8 Si$_x$Ge$_{1-x}$ alloys. Red crosses represent symmetrically independent site occupancy configurations. Blue diamonds represent special quasi-random structures (SQS).}\label{fig:enth_form}
\end{figure}

Vegard's law \citep{den91} is conventionally employed to empirically approximate the relationship between a specific property and the composition of a binary alloy AB. It is given in a simple linear form as $a_x = (1-x)a^A + a^B - bx(1-x)$, where $a_x$ is the parameter of the alloy, $a^A$ and $a^B$ the parameters of the pure phases and $b$ a bowing parameter that quantifies the deviation from a linear relationship. The lattice parameter of ST12 and BC8 alloys obtained after relaxation are shown in Figure \ref{fig:lattice}. We give two predictions for the alloy lattice parameters, the first one is a weighted average according to the symmetry derived probability\cite{gra07} of all PC configurations for a particular composition $x$, and the second one is the direct result of the SQS relaxation. Both predictions are in excellent agreement and approximately follow Vegard's law with just a slight bending (bowing parameters $b_{BC8} = 0.03774$ and $b_{ST12} = 0.04307$). Furthermore, Figure \ref{fig:enth_form} shows the enthalpy of formation for the entire compositional range (including SQS as blue diamonds), given as 
\begin{equation}
\label{eq:1}
\Delta H = H_{\text{Si}_x\text{Ge}_{1-x}} - (1-x)H_{\text{Ge}} - xH_{\text{Si}},
\end{equation}
where $H_{\text{Si}_x\text{Ge}_{1-x}}$ is the total enthalpy of the alloy, $H_\text{Ge}$ and $H_\text{Si}$ the enthalpies of the respective end-member compositions. Minima in the hull would indicate possible intermediate ground states, however, for both BC8 and ST12, no particular configurations more stable than the pure phases were identified. Nevertheless, it can be observed that the formation enthalpies of SQS supercells fall right within the spread of all PC configurations, indicating that the SQS are indeed good representations of each ensemble.

\begin{figure}
\includegraphics[width=0.5\textwidth]{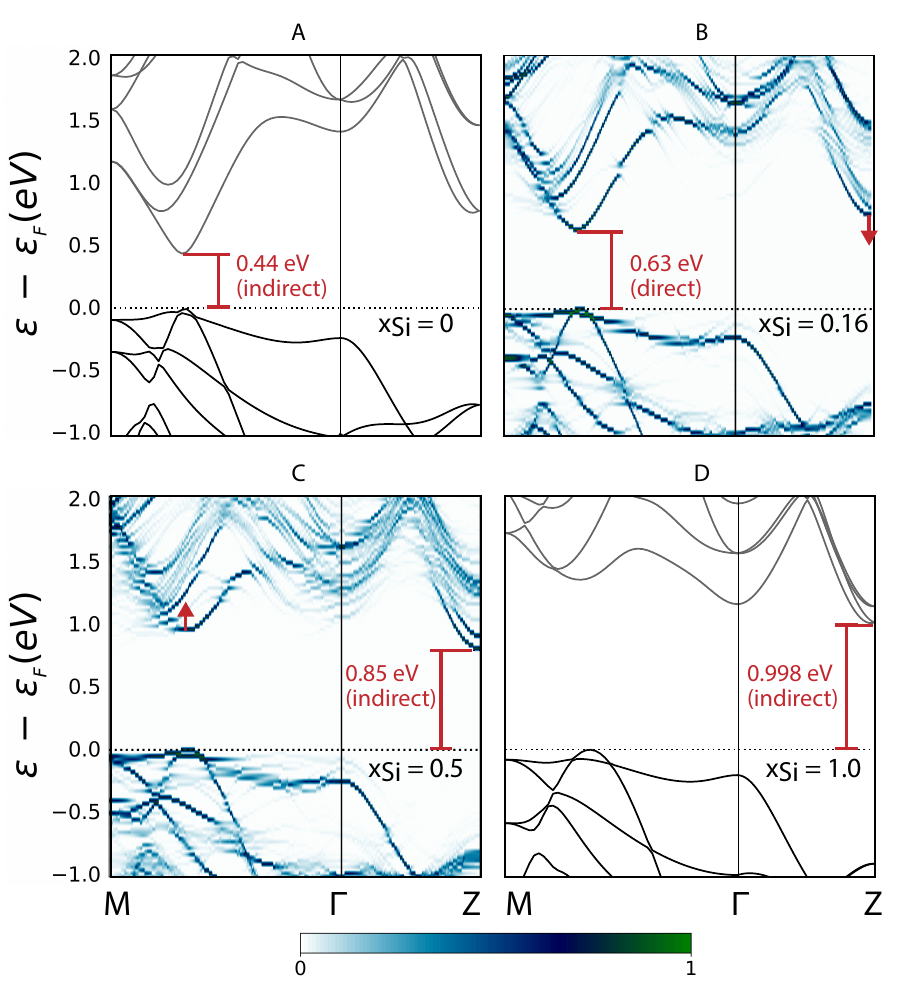}
\caption{ST12 alloy band structure plots in the region $M-\Gamma-Z$. (A) and (D) are plots derived from the primitive end-member simulation cells, (B) and (C) show unfolded effective band structures derived from special quasi-random structures (SQS). (A)~Pure ST12 Ge with a well studied indirect band gap. (B)~Effective SQS band structure for $x=0.166$ showing a direct fundamental band gap between $\Gamma$ and $M$. (C)~Effective SQS band structure for $x=0.5$ where the original CBM has shifted to $Z$, resulting in a large indirect band gap. (D)~Band structure of pure hypothetical ST12 Si.}\label{fig:gaps}
\end{figure}

To give a meaningful description of the band structure of intermediate compositions, we unfolded the SQS bands using the BandUP code \cite{med14}. Then we evaluated the band gap energy as well as the CBM/VBM positions using the SUMO code \cite{gan18}. For the ST12 structure we found that increasing Si content in Ge ST12, gradually opens the gap while at the same time, the CBM and VBM are slightly shifted, narrowing the already small difference between direct and indirect transition. This culminates in a direct band gap at x$_\text{Si} \approx 0.16$ as shown in Figure \ref{fig:gaps}. To confirm this observation, we repeated the band structure calculation for this composition using a total of 1000 K-points in the region $M-\Gamma-Z$ with the same result. Further introduction of Si into ST12 Ge led to the retraction of the original CBM along the $\Gamma-M$ line and the formation of a new CBM at $Z$, resulting in an indirect band gap (0.85~eV at $x_\text{Si}=0.5$) steadily widening until the end-member composition was reached. For the ST12 alloys, SQS band calculations were not repeated in the HSE06 formalism as they are computationally extremely demanding. However, a qualitatively similar behavior as described above for ST12 end-member compositions is expected.\\
For all BC8 alloys, PBEsol calculations predicted a metallic nature throughout the entire compositional range. Simulations using the same HSE06 parameters as for the end-member compositions also resulted in indirect band gaps for all intermediate compositions. Even the smallest addition of Ge that we modeled ($x_\text{Si} = 0.94$) resulted in an indirect band gap. Increasing the Ge content did not change the overall band structure features all the way up to end-member Ge.

In summary, we have shown that it should be possible to {\it compositionally tune} the band gap in ST12 Si$_x$Ge$_{1-x}$ alloys to become direct at low Si concentrations ($x_\text{Si}\approx0.16$). Given that samples with such composition have already been synthesized following a well established method\citep{ser14}, it should be possible to validate our predictions by further experiments. For the BC8 structure, we have demonstrated that widening the narrow band gap of pure BC8 Si by alloying it with Ge is not possible. In this case, more traditional avenues for band gap tuning (e.g. other dopants, inducing strain, etching) may be more rewarding. From our overall results, we conclude that Si$_x$Ge$_{1-x}$ ST12 alloys with $0.1 \le x_\text{Si} \le 0.2$ are viable candidates for direct band gap materials based to a considerable percentage on commonly available Si. Therefore, we suggest further experimental studies on these alloys to narrow down the compositonal range and confirm our findings.

\begin{acknowledgments}
The authors gratefully acknowledge the Gauss Centre for Supercomputing e.V. (www.gauss-centre.eu) for funding this project by providing computing time through the John von Neumann Institute for Computing (NIC) on the GCS Supercomputer JUWELS at J\"ulich Supercomputing Centre (JSC) under project {\bf abinitiomodmatsgeo}. JW and MNV were supported by the Helmholtz Association through \textit{funding of first-time professorial appointments of excellent women scientists (W2/W3).} The authors are thankful to Hans Josef Reichmann and Monika Koch-M\"uller for helpful discussions during the writing of this manuscript.
\end{acknowledgments}

\appendix

\bibliography{sige}

\end{document}